\documentclass[aps,
prd,
english,
nofootinbib,
longbibliography,
showkeys,
reprint
]{revtex4-2}

\pdfoutput=1
\usepackage{amsfonts,amsmath,amssymb}
\usepackage{mathrsfs}
\usepackage{amsthm}
\usepackage{graphicx}
\usepackage[utf8]{inputenc}
\usepackage{hyperref}
\usepackage{babel}
\usepackage{listings}
\usepackage{matlab-prettifier}
\usepackage{enumitem}
\usepackage{lipsum}
\usepackage{tikz}
\usetikzlibrary{arrows.meta,calc,decorations.pathreplacing}

\newcommand\be{\begin{equation}}
\newcommand\ee{\end{equation}}
\newcommand\bea{\begin{eqnarray}}
\newcommand\eea{\end{eqnarray}}

\begin{document}

\def\rhoo{\rho_{_0}\!} 
\def\rhooo{\rho_{_{0,0}}\!} 
\def\G{{\widetilde\Gamma}}

\newcommand{\dd}{\,d}
\newcommand{\zh}{\hat z}

\newcommand{\PP}{\operatorname*{PP}}
\newcommand{\Res}{\operatorname*{Res}}
\newcommand{\cL}{\mathcal L}
\newcommand{\oL}{\mathscr L}
\newcommand{\cQ}{\mathcal Q}
\newcommand{\cW}{\mathcal W}
\newcommand{\cV}{\mathcal V}
\newcommand{\om}{\omega}
\newcommand{\B}{\widehat B}
\newcommand{\hV}{\widehat V}

\begin{flushright}
\phantom{
{\tt arXiv:2006.$\_\_\_\_$}
}
\end{flushright}

{\flushleft\vskip-1.4cm\vbox{\includegraphics[width=1.15in]{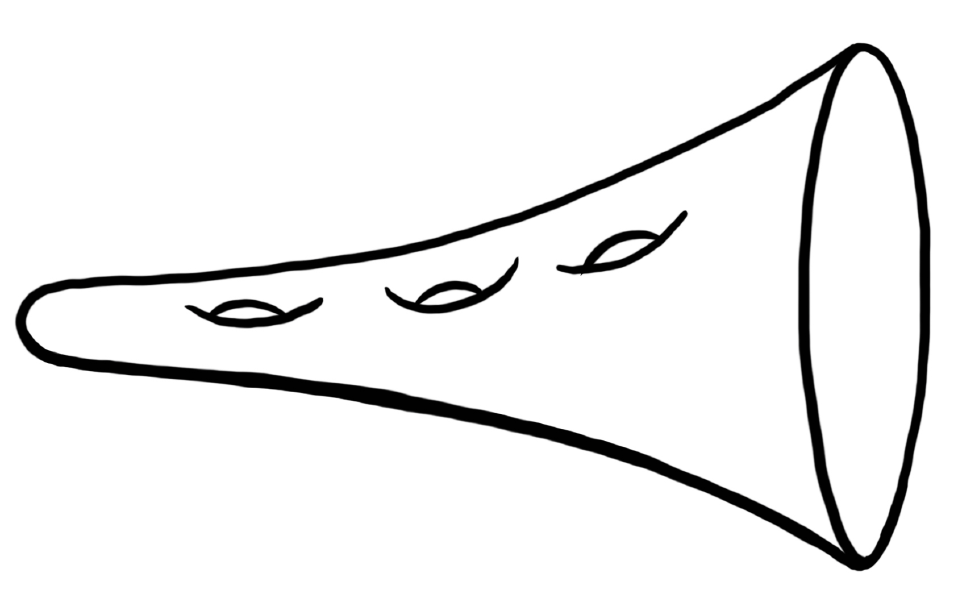}}}

\title{
  {${\cal N}{=}1$ Supersymmetry, Weil-Petersson Volume Recursion, and a Spectral Curve 
  }}
\author{Clifford V. Johnson}
\email{cliffordjohnson@ucsb.edu}

\affiliation{Department of Physics, Broida Hall,   University of California, 
Santa Barbara, CA 93106, U.S.A.}


\begin{abstract}

The Stanford-Witten-Norbury generalization of Mirzakhani's volume recursion   computes $V^{(2m)}_{g,n}(\{b_i\})$, the  Weil-Petersson volumes of the  moduli space of  ${\cal N}{=}1$ supersymmetric Riemann surfaces of  genus~$g$ with $n$ Neveu-Schwarz boundaries of geodesic lengths~$b_i$ ($i{=}1,\ldots,n$), and $2m$ Ramond punctures.  Recently, a spectral curve has been derived  that allows their Laplace transforms  $W^{(2m)}_{g,n}(\{{\hat z}_i\})$ to be computed using  topological recursion. We prove that the Stanford-Witten-Norbury volume recursion is directly derivable from the spectral curve. An alternative volume recursion can also be derived from it. The difference comes from whether the Ramond information is in the initial data, or in the volume recursion's kernels. The latter invites a geometrical understanding.

\end{abstract}


\maketitle

\section{The Stanford-Witten-Norbury volume recursion}
We will use $\hV_{g,n}(s;b_1,\ldots,b_n)$ to denote 
 Norbury's~\cite{norbury2024superweilpeterssonmeasuresmoduli} Ramond-deformed volume generating function, which has an expansion in even powers of $s$:
\begin{equation}
  \hV_{g,n}(s;b_1,\ldots,b_n)
  = \sum_{m=0}^\infty 
  s^{2m}\,
  \hV_{g,n}^{(2m)}(b_1,\ldots,b_n)\ ,
  \label{eq:Vhat-def}
\end{equation}
 where $\hV_{g,n}^{(2m)}$ is the (gas-normalized)\footnote{There is a different volume normalization involving pulling out an additional$\frac{
1}{(2m)!}$ factor from ${\widehat V}^{(2m)}_{g,n}$~\cite{norbury2024superweilpeterssonmeasuresmoduli}. We won't need it here.} volume  with $2m$ Ramond (R) punctures,
and  $n$ is the number of Neveu-Schwarz (NS) boundaries with lengths~$b_i$, ($i{=}1,\cdots,n$).
It is also useful to have  a compact notation for  labels to be summed over, so
write $I,J,K$, for sets of indices, {\it e.g.}:
\begin{equation}
  K=\{2,\ldots,n\},
  \qquad
  b_I=(b_i)_{i\in I}\ .
  \label{eq:K-I-notation}
\end{equation}
The volume recursion can then  written in the form~\cite{Stanford:2019vob,Norbury:2020vyi,norbury2024superweilpeterssonmeasuresmoduli}:

\begin{widetext}
    
\begin{align}
  b_1\hV_{g,n}(s;b_1,b_K)
  ={}&\frac12\int_0^\infty\!\int_0^\infty
  x y\,D(b_1,x,y)
  \Bigg[
  \hV_{g-1,n+1}(s;x,y,b_K)
  +\sum_{\substack{g_1+g_2=g\\ I\sqcup J=K}}
  \hV_{g_1,|I|+1}(s;x,b_I)\,
  \hV_{g_2,|J|+1}(s;y,b_J)
  \Bigg]\,\dd x\,\dd y
  \nonumber\\
  &\hspace{3.0cm}+\sum_{j=2}^n\int_0^\infty
  x\,R(b_1,b_j,x)\,
  \hV_{g,n-1}(s;x,b_{K\setminus\{j\}})\,\dd x\ ,
  \label{eq:Norbury-volume-recursion}
\end{align}
\end{widetext}
where the sum over \(I\sqcup J=K\) is over ordered decompositions, and kernels $D$ and $R$ will be defined shortly.
As a
formal recursion in powers of \(s^2\), \eqref{eq:Norbury-volume-recursion} is solved order
by order. As written, it is understood that
\(\hV_{g,n}{=}0\) for~\(g{<}0\).

Crucially, terms in the sums containing a disc or cylinder
are {\it not} discarded (as they are in the reduction to the purely NS form of the recursion~\cite{Stanford:2019vob}, where such terms are defined to vanish). These  non-zero ``unstable'' insertions are responsible for introducing the Ramond data into all higher volumes.  
They can be described by specifying their Laplace transforms: 
\begin{equation}
\widehat F({\hat z},s) = \cL_b[\hV_{0,1}(s;b)]({\hat z})\ ,
  \label{eq:unstable-disc-input}
\end{equation}
and
\begin{equation}
\widehat C({\hat z}_1,{\hat z}_2;s) = \cL_{b_1,b_2}[\hV_{0,2}(s;b_1,b_2)]({\hat z}_1,{\hat z}_2)\ ,
  \label{eq:unstable-cylinder-input}
\end{equation}
where we  define the Laplace transform on volumes  by:
\begin{equation}
  \cL[f]({\hat z}_1,\ldots,{\hat z}_n)
  =\int_0^\infty\prod_{i=1}^n b_i\,\dd b_i\,
  f(b_1,\ldots,b_n)e^{-\sum_i b_i{\hat z}_i}.
  \label{eq:laplace-convention}
\end{equation}
(We will use $\oL[f]$ for the ordinary Laplace transform.)
The function \(\widehat F\) is determined by a disc equation reviewed in Section~\ref{sec:disc-proof}, and the cylinder equation determining  $\widehat C$ (in terms of $\widehat F$) will be derived in Section~\ref{sec:cylinder-proof}. The first few terms of each (as a series in $s^2$) are:
\begin{eqnarray}
   \widehat F({\hat z_1},s)=\frac{s^2}{2\hat z_1^2}+\cdots\ ,\quad 
    \widehat C({\hat z}_1,{\hat z}_2;s) =
    \frac{s^2}{2\hat z_1^2\hat z_2^2}
    +\cdots
\end{eqnarray}
and so they vanish as $s\to0$, the purely NS case.

The two kernels appearing in \eqref{eq:Norbury-volume-recursion} are the usual ones that first appeared in Stanford and Witten~\cite{Stanford:2019vob} (and unpublished work by Norbury) for the purely NS case.  In other words, they (crucially) have no Ramond data within them. They are:
\begin{equation}
  D(b_1,x,y)=\frac1{4\pi}\left[
  \frac1{\cosh\left(\frac{b_1-x-y}{4}\right)}
  -\frac1{\cosh\left(\frac{b_1+x+y}{4}\right)}
  \right],
  \label{eq:D-kernel-standard-statement}
\end{equation}
and
\begin{equation}
  R(b_1,b_2,x)=\frac12\left[D(b_1+b_2,x,0)+D(b_1-b_2,x,0)\right].
  \label{eq:R-kernel-standard-statement}
\end{equation}
Just as for the original bosonic version of such recursion, due to Mirzakhani~\cite{Mirzakhani:2006fta},
the kernel $D$ connects the outer boundary $b_1$ to two inner geodesic boundaries with lengths $(x,y)$ (which are integrated over) and so encodes a kind of pair-of-pants object (or three-holed sphere). The kernel $R$ (also associated with  a pair-of-pants) connects the outer boundary $b_1$ to the marked boundary~$b_j$, with $x$ the length of an internal boundary that is integrated over. $D$ and $R$ together  contain precise geometrical information about the relative weights of classes of geodesics that leave the $b_1$-boundary  orthogonally and  either return to the boundary, self-intersect, or reach the other boundary ($b_j$, when present). Ref.~\cite{Stanford:2019vob} showed that there were suitable analogues of these objects to the ${\cal N}{=}1$ supersymmetric case under discussion, and derived the above expressions for them.

\section{The Spectral Curve}
On the other hand, there has been recent progress in swiftly computing the $V^{(2m)}_{g,n}(\{b_i\})$ directly using random matrix model techniques~\cite{Johnson:2026twg,Johnson:2026jls}. Their Laplace transforms, $W^{(2m)}_{g,n}(\{\zh_i\})$ can be built from data obtained by perturbatively solving a certain  ordinary differential equation called a ``string equation''. These matters won't concern us here, but a key output of ref.~\cite{Johnson:2026jls} was a new spectral curve, made by analytically continuing the leading spectral density of a random matrix model. It is:
\begin{widetext}

    \begin{equation}
\rho_0(E)=
\frac{1}{2\pi \hbar}
\left[
\int_{E_0}^{E}
\frac{\pi I_1(2\pi\sqrt {u_0})}{\sqrt {u_0}\sqrt{E-u_0}}\,du_0
+
I_0(2\pi\sqrt{E_0})\frac{\sqrt{E-E_0}}{E}
\right]\ ,\qquad E>E_0\ ,
\label{eq:exact-density-i0}
\end{equation}
and moreover there is a parameter $\G$ related to threshold energy $E_0$ by:
\begin{equation}
\G=\sqrt{E_0}\,I_0(2\pi\sqrt{E_0})\ ,\quad\text{giving}\quad
\label{eq:endpoint-equation}
  E_0=\G^2-2\pi^2\G^4+\frac{13\pi^4}{2}\G^6-\frac{230\pi^6}{9}\G^8+\cdots .
\end{equation}
Identifying the ``hard-edge'' coordinate by $E=-\hat z^2$,   the spectral curve is defined as:

\begin{equation}
  X({\hat z})={\hat z}^2\ ,
  \qquad
  Y_{\G}({\hat z})=-\pi \hbar i\,\rho_0(-{\hat z}^2)\ .
  \label{eq:hard-edge-curve}
\end{equation}
Topological recursion~\cite{Chekhov:2006vd,Eynard:2007kz} can be applied to this curve to extract the $W^{(2)}_{g,n}$ in yet another way. However, it needs the additional data of a Bergman kernel. Crucially  it is {\it not} the commonly used  Cauchy kernel in \({\hat z}\), but instead:\footnote{In ref.~\cite{Johnson:2026jls} this was denoted $\B (\zh_1,\zh_2)$, but here we will  drop that to avoid certain potential notational confusions later.}
\begin{eqnarray}
  B_{\G}({\hat z}_1,{\hat z}_2)
  =\frac{{\hat z}_1{\hat z}_2}{z_1z_2}\,
  \frac{\dd{\hat z}_1\dd{\hat z}_2}{(z_1-z_2)^2}\ ,
\quad
\text{with}\quad
z_i&=&\sqrt{{\hat z}_i^{\,2}+E_0} \ ,   
  \label{eq:pull-back-Bergman}
\end{eqnarray}
simply the pull-back of the Cauchy Bergman in soft-edge coordinates $z_i$ to the kernel in hard-edge coordinates~\cite{Johnson:2026jls}.

This set, $\{X(\zh),Y_\G(\zh); B_\G\}$,  defines the starting data for the topological recursion, with $\omega_{0,1}=X(\zh)dY(\zh)$ and $\omega_{0,2}=B_{\G}({\hat z}_1,{\hat z}_2)$. The recursion is:
    \begin{align}
 \omega_{g,n}({\hat z}_1,{\hat z}_S)
 ={}&\operatorname*{Res}_{{\hat z}=0} K({\hat z}_1,{\hat z})
 \Bigg[
 \omega_{g-1,n+1}({\hat z},\sigma({\hat z}),{\hat z}_S)
+\sum_{\substack{g_1+g_2=g\\ I\sqcup J=S}}^{\prime}
 \omega_{g_1,|I|+1}({\hat z},{\hat z}_I)
 \omega_{g_2,|J|+1}(\sigma({\hat z}),{\hat z}_J)
 \Bigg].
 \label{eq:tr-recursion}
\end{align}
    and the prime on the sum means that the unstable terms involving $\omega_{0,1}$ are omitted. 
 \end{widetext}The recursion kernel is: \begin{equation}
 K({\hat z}_1,{\hat z})=
 -\frac{\displaystyle\int_{\sigma({\hat z})}^{{\hat z}}B_\G({\hat z}_1,\zeta)}
 {2\bigl(Y_\G({\hat z})-Y_\G(\sigma({\hat z}))\bigr)dX({\hat z})}\ ,
 \label{eq:kernel-def}
\end{equation}
where at the ramification point $\zh=0$ there is the involution $\sigma(\zh)=-\zh$.

Once the $\omega_{g,n}$ are obtained by running topological recursion on this curve, the $W_{g,n}$s are defined by:\footnote{In fact, the $W_{g,n}$ can be readily computed in terms of the matrix model function $u_0(x)$ and its derivatives using the formalism of refs.~\cite{Johnson:2026twg,Johnson:2026jls}, without the need for topological recursion. However the focus on this paper is on the topological recursion approach to the spectral curve and its relation to volume recursion.}
\begin{equation}
 \omega_{g,n}(\zh_1,\ldots,\zh_n)
 =W_{g,n}(\zh_1,\ldots,\zh_n)\prod_{i=1}^{n}d\zh_i.
 \label{eq:omega-to-w}
\end{equation}
They have an expansion in $\G^2$:
\begin{equation}
  W_{g,n}(\G;\zh_1,\ldots,\zh_n) = \sum_{m=0}^\infty W_{g,n}^{(2m)}(\zh_1,\ldots,\zh_n)\G^{2m}\ ,
  \label{eq:W-expansion}
\end{equation}
 They are simply the Laplace transforms of the following volumes, {\it i.e.}:
\begin{equation}
    W_{g,n}^{(2m)}(\zh_1,\ldots,\zh_n) = {\cal L}[V^{(2m)}_{g,n}](\zh_1,\ldots,\zh_n)\ ,
\end{equation}

In comparing ref.~\cite{Johnson:2026jls}'s  volume conventions to Norbury's, we are using  hats for volumes~({\ref{eq:Vhat-def}}), of ref.~\cite{norbury2024superweilpeterssonmeasuresmoduli}, and no hats for ref.~\cite{Johnson:2026jls}'s,
with translation:
\begin{equation}
  V^{(2m)}_{g,n}(b_1,\ldots,b_n)
  =(-1)^{n}\,
  \hV^{(2m)}_{g,n}(b_1,\ldots,b_n)\ ,
  \label{eq:Johnson-Norbury-coefficient-map}
\end{equation}
The R-puncture counting variable used by ref.~\cite{norbury2024superweilpeterssonmeasuresmoduli} is denoted \(s\) and  its relation to $\G$ is:
\begin{equation}
  s^2=-\G^2 .
  \label{eq:s-Gamma}
\end{equation} 
Just as with the volume recursion, one performs the topological recursion  order by order in a $\G^2$ expansion. Order~$\G^0$ is just the ordinary case with only NS boundaries~\cite{Stanford:2019vob}, as can be seen by setting $\G{=}0$ whence $E_0{=}0$ and the definitions above give the undeformed data~\cite{Norbury:2017eih}:
\begin{equation}
  Y_0({\hat z})=-\frac{\cos(2\pi{\hat z})}{2{\hat z}},
  \qquad
  B_0({\hat z}_1,{\hat z}_2)=\frac{\dd{\hat z}_1\dd{\hat z}_2}{({\hat z}_1-{\hat z}_2)^2}\ ,
  \label{eq:undeformed-data}
\end{equation}
from which we can construct the undeformed kernel:
\begin{equation}
  K_0({\hat z}_1,{\hat z})
  =\frac{{\hat z}\,\dd{\hat z}_1}{2\cos(2\pi{\hat z})({\hat z}_1^2-{\hat z}^2)\dd{\hat z}}\ .
  \label{eq:K-undeformed}
\end{equation}
Even with R-punctures present, on the volume recursion side, ref.~\cite{norbury2024superweilpeterssonmeasuresmoduli} noted that the kernels do not involve $s$, and so are the same kernels as used in the pure NS case. So the first step is to recall how the  kernels come from the order zero spectral curve data, in particular the topological recursion kernel (\ref{eq:K-undeformed}).

\begin{widetext}

\section{The origin of the volume recursion kernels}
\label{sec:kernel-origin}

It is prudent to rewrite the  undeformed topological recursion kernel as:
\begin{equation}
  K_0({\hat z}_1,{\hat z})
  =\frac{1}{2Q_Y(\zh)}
  \left(\frac1{{\hat z}_1-{\hat z}}-\frac1{{\hat z}_1+{\hat z}}\right)
  \frac{\dd{\hat z}_1}{\dd{\hat z}}\ ,\quad \text{where}\quad 
    Q_Y({\hat z}) := 2{\hat z}\,Y(-{\hat z}^{2}) .
  \label{eq:cauchy-partial-fraction}
\end{equation}

\end{widetext}
 In this undeformed \(N{=}1\) case  we have 
   $Q_Y({\hat z})=\cos(2\pi {\hat z})$, 
 but we can leave things a bit more general for the next few steps,
and try to determine (in the spirit of refs.~\cite{Eynard:2007fi,Stanford:2019vob}) how it would originate from a Laplace transform of a pair-of-pants with external boundary length $b_1$ (dual to variable ${\hat z}_1$) and internal boundary lengths $x$ and $y$.

Upon inverse Laplace transforming in the  ${\hat z}_1$ variable, the partial fractions
contribute factors according to:
\begin{equation}
  \frac1{{\hat z}_1-{\hat z}}\longleftrightarrow e^{+{\hat z} b_1}\ ,
  \qquad
  \frac1{{\hat z}_1+{\hat z}}\longleftrightarrow e^{-{\hat z} b_1}\ .
  \label{eq:cauchy-exponentials}
\end{equation}
The two internal boundaries contribute the Laplace weights
\(e^{-{\hat z}x}e^{-{\hat z}y}=e^{-{\hat z}(x+y)}\), and so the two  terms produce the shifted arguments:
\begin{equation}
  e^{\pm{\hat z} b_1}e^{-{\hat z}(x+y)}=e^{-{\hat z}(x+y\pm b_1)}\ .
  \label{eq:shift-origin}
\end{equation}
So we can represent the structure of what we see in $K_0(\zh_1,\zh)$ including the 
\(1/Q_Y\) factor (which eventually will be evaluated at $\zh_1$) as:
\begin{equation}
  \frac{1}{2}D_Y(x+y,b_1)
  =
  \int_{-i\infty}^{i\infty}\frac{\dd {\hat z}}{2\pi i}\,
  \frac{e^{-(x+y){\hat z}}}{Q_Y({\hat z})}\,\sinh(b_1{\hat z})\ .
  \label{eq:general-D-kernel-inverse}
\end{equation}
since the factor \(\sinh(b\zh)\) comes the sum of  the pair of
Cauchy-originated factors
$  \sinh(b_1\zh){=}\frac12\left(e^{b_1\zh}-e^{-b_1\zh}\right)$.
 Laplace transforming back in \(b_1\), with \({\hat z}_1\) the external
hard-edge transform variable, gives:
 \begin{equation}
   \int_0^\infty e^{-{\hat z}_1b_1}\sinh(b_1\zh)\dd b_1
   =\frac{\zh}{{\hat z}_1^{2}-\zh^{2}}=\sum_{k=0}^\infty\frac{\zh^{2k+1}}{{\hat z}_1^{2k+2}}\ .
   \label{eq:sinh-hard-edge-cauchy}
 \end{equation}
We can now  see how the kernel works when acting on something (in length space). From the recursion~(\ref{eq:Norbury-volume-recursion}) we see that on some (odd) function $P(x,y)$ we have actions of the form: 
\begin{eqnarray}
\label{eq:D-action}
&&\hskip-0.5cm\int_0^\infty\!
\int_0^\infty \frac12D_Y(x+y,b_1)P(x,y)dxdy\\
&&=\int \frac{d\hat z}{2\pi i}\frac{\sinh(b_1\hat z)}{Q_Y(\hat z)}\int_0^\infty\!\int_0^\infty e^{-\hat z x-\hat z y}P(x,y)dxdy\ ,\nonumber
\end{eqnarray}where we used~(\ref{eq:general-D-kernel-inverse}). Notice that
 the inner double integral is 
$ {\oL}[P](\hat z,\hat z)$, which, given that $P(x,y)$ is made from odd polynomials in $x$ and $y$, is now made from inverse powers of $\zh$. Laplace transforming on $b_1$ and using~(\ref{eq:sinh-hard-edge-cauchy}) will (upon doing the $\zh$ integral), simply result in  the residue of the product of $ {\oL}[P](\hat z,\hat z)/Q_Y(\zh)$ with the sum in (\ref{eq:sinh-hard-edge-cauchy}). This selects the polar part (the Laurent expansion in inverse powers of~$\zh$) of $ {\oL}[P](\hat z,\hat z)/Q_Y(\zh)$, evaluated at $\zh=\zh_1$.

There is one important subtlety: The residue is a {\it half}-residue since the pole lies exactly on the contour (requiring the usual half-circle detour), so there is a factor of a half in front of the polar part, conveniently cancelling with the half in front of $D_Y$ on the left hand side of~\eqref{eq:D-action}.

Let us record our result (derived in ref.~\cite{Norbury:2020vyi} using a slightly different approach)  which  will be useful later:
\begin{widetext}

\begin{equation}
  \oL_{b_1}\left[
  \int_0^\infty\!\int_0^\infty
  D_Y(b_1,x,y)P(x,y)\dd x\dd y
  \right]({\hat z}_1)
  =
  \PP_{{\hat z}_1=0}\left[
   \frac{\oL[P]({\hat z}_1,{\hat z}_1)}{Q_Y({\hat z}_1)}
   \right]
\longrightarrow
  \PP_{{\hat z}_1=0}\left[
  \frac{\oL[P]({\hat z}_1,{\hat z}_1)}{\cos(2\pi {\hat z}_1)}
  \right]\ ,
  \label{eq:D-transform-identity-from-kernel-origin}
\end{equation}
    \end{widetext}
 where the last step  recalls that we have $Q_Y(\zh_1){=}\cos(2\pi\zh_1)$, the case of most interest here.

For writing $D$ explicitly, the  following result is useful:
%
\begin{equation}
  F_0(u)
  =
  \int_{-i\infty}^{i\infty}
  \frac{\dd{\hat z}}{2\pi i}\,
  \frac{e^{u{\hat z}}}{\cos(2\pi{\hat z})}
  =
  \frac{1}{4\pi\cosh(u/4)}\ ,
  \label{eq:sech-inverse-transform}
\end{equation}
and~(\ref{eq:general-D-kernel-inverse}) gives (dropping the $Y$ subscript on $D_Y$):
\begin{equation}
  D(b_1,x,y)=
  F_0(b_1-x-y)-F_0(b_1+x+y)
  \ ,
  \label{eq:D-kernel}
\end{equation}
yielding precisely the kernel~\eqref{eq:D-kernel-standard-statement}.  

The second kernel in the volume recursion, \(R\), is built from  $D$ according to~(\ref{eq:R-kernel-standard-statement}).  
%
%
Analogous steps to what we did here for $D$ will allow us to deduce the counterpart of~(\ref{eq:D-transform-identity-from-kernel-origin}) for~$R$'s action. We shall leave that for Section~\ref{sec:cylinder-proof}.

\section{A choice of Volume Recursion}
\label{sec:alternatives}
When $\G$ (hence $s$) is turned back on, bringing the R-punctures back into the story, we  have a choice. The first is to inverse Laplace transform the full kernel~(\ref{eq:kernel-def}), giving a version of what we've just done with $K$ and $R$ replaced by $\G$-- (hence $s$--) dependent kernels.  This can be explored order by order. For example, at order $\G^2$ we can expand the kernel~(\ref{eq:kernel-def}) to get:
\begin{equation}
  K_{\G}(\zh_1,\zh)=K_0(\zh_1,\zh)
  +\G^2 K_2(\zh_1,\zh)+O(\G^4)\ ,
  \label{eq:KG-expansion}
\end{equation}
and the order $\G^2$ piece is:
\begin{widetext}
\begin{eqnarray}
  &&\hskip-0.5cmK_2(\zh_1,\zh)=\frac{\dd\zh_1}{\dd\zh}
  \left[
  \frac{1}{4\cos(2\pi\zh)\zh_1^2\zh}-\frac{1}{4\cos(2\pi\zh)^2\zh(\zh_1^2-\zh^2)}
  \right]\ ,\nonumber
  \label{eq:K2-explicit}
\end{eqnarray} where the first term comes from the corrections to the Bergman kernel $B_{\G}$ away from the Cauchy form and the second term originates from the leading deformation of the spectral curve $Y(\zh)$ away  from the NS case.

After some Laplace transform steps that mimic the work of the previous section, it is possible to compute how the length space volume kernel gets corrected by the Ramond data:
\begin{equation}
    D_{\G}(b_1,x,y)=D_0(b_1,x,y)+\G^2 D_2(b_1,x,y)+\cdots
\end{equation}
where $D_0(b_1,x,y)$ is given in~(\ref{eq:D-kernel-standard-statement}).
Interestingly, $D_2$ has two very different kinds of contribution, $D^{(B)}_2$ and $D^{(Y)}_2$, traceable to the form of the contributions in $K_2$ above. The second term still admits the kind of  partial fraction decomposition we saw before in~\eqref{eq:cauchy-partial-fraction}, so it results in 
a contribution  $D^{(Y)}_2$ that is again a difference of two terms built from the same function evaluated at  shifted arguments $b+(x+y)$ or $b-(x+y)$:
\begin{equation}
    D^{(Y)}_{2}(b,x,y)
=
-\frac18 F_2^{+}(b-x-y)
+
\frac18 F_2^{-}(b+x+y)\ ,
\quad\text{
where
}
F_2^{\pm}(u)
=
\int_{-i\infty}^{i\infty}
\frac{d\hat z}{2\pi i}
\frac{e^{\pm\hat z u}}{\cos(2\pi\hat z)^2\hat z^2}\ ,
\end{equation} but the integral defining $F_2^\pm$ yields a cumbersome final form which we will not write.\footnote{It is apparently expressible in terms of dilogarithms and trilogarithms, or 2nd and 3rd order Legendre $\chi$-functions, according to taste.} A very new feature comes from the first term in $K_2$, which due to the lack of that partial fraction form does not yield the same kind of structure. Instead the simple $\zh_1^{-2}$ yields $b_1$ after the transform, and the remaining integral over a single power of $\cos(2\pi\zh)\zh$ is solvable, with the result that the deformed $B$ contribution to $D_2(b_1,x,y)$ is:
\begin{equation}
    D^{(B)}_{2}(b_1,x,y)
=
b_1
\int_{-i\infty}^{i\infty}
\frac{d\hat z}{2\pi i}
\frac{e^{-\hat z(x+y)}}{4\hat z\cos(2\pi\hat z)}=\frac{b_1}{2\pi}
\arctan\left(e^{-(x+y)/4}\right)\ ,
\end{equation}
which was striking enough to display.

The main point here is that while this clearly will still give a recursion scheme, and while it would be interesting to understand if there is some geometry being revealed by the Ramond-deformed kernel ($D$, and the $K$ derived from it), the recursion itself will rapidly gets unwieldy due to the complexity of $D$ as we go to higher orders in $\G^2$.

There is another choice, which is to find a way to keep  the Ramond data out of the kernels and instead put it into the initial data. This is the route we will explore below, not the least since a primary goal is to demonstrate the equivalence between the spectral curve data presented above and the Stanford-Witten-Norbury scheme, where the Ramond data are indeed entirely in the initial data. This means that there should be a way of re-presenting the recursion above to allow that.

\section{Rearranging recursion}

\noindent Let's rewrite ref.~\cite{Johnson:2026jls}'s full topological recursion in the following form:
\begin{equation}
  \om^{\G}_{g,n}({\hat z}_1,{\hat z}_I)
  =\Res_{{\hat z}=0} K_{\G}({\hat z}_1,{\hat z})\,
  \cQ^{\G}_{g,n}({\hat z},-{\hat z};{\hat z}_I)\ ,
  \label{eq:TR-full}
\end{equation}
where \(I=\{2,\ldots,n\}\), and:
\begin{align}
  \cQ^{\G}_{g,n}({\hat z},-{\hat z};{\hat z}_I)
  ={}&\om^{\G}_{g-1,n+1}({\hat z},-{\hat z},{\hat z}_I)
+\sum_{\substack{g_1+g_2=g\\J\sqcup J'=I}}^{\prime}
  \om^{\G}_{g_1,|J|+1}({\hat z},{\hat z}_J)
  \om^{\G}_{g_2,|J'|+1}(-{\hat z},{\hat z}_{J'})\ ,
  \label{eq:Q-def}
\end{align}
with \(\om^{\G}_{0,2}=B_{\G}\).  The prime omits  unstable disc factors in the usual topological recursion convention.
\end{widetext}
Before the step involving the residue, locally the recursion is governed by:
\begin{equation}
  2\om^{\G}_{0,1}({\hat z})\,\om^{\G}_{g,n}({\hat z},{\hat z}_I)
  =\cQ^{\G}_{g,n}({\hat z},-{\hat z};{\hat z}_I)
  \ ,
  \label{eq:loop-equation-full}
\end{equation}
up to  terms that are regular at the branch point.    This is because $\om^{\G}_{0,1}({\hat z})d\zh =X(\zh)dY(\zh)$, and the kernel can be rewritten as:
\begin{equation}
  K_{\G}(z_1,z)
  =\frac12\frac{\displaystyle\int_{\sigma(z)}^z B_{\G}(z_1,\cdot)}
  {\om^{\G}_{0,1}(\sigma(z))-\om^{\G}_{0,1}(z)},
  \qquad \sigma({\hat z})=-{\hat z}\ .
  \label{eq:EO-kernel-abstract}
\end{equation}
Now  split the unstable data as:
\begin{equation}
  \om^{\G}_{0,1}=\om^0_{0,1}+\Delta\om_{0,1}\ ,
  \qquad
  \om^{\G}_{0,2}=\om^0_{0,2}+\Delta\om_{0,2}\ ,
  \label{eq:unstable-split}
\end{equation}
where the 0~superscript means the Ramond-independent parts. Moving the disc correction to the right gives:
\begin{equation}
  2\om^0_{0,1}\,\om^{\G}_{g,n}
  =\cQ^{\G}_{g,n}-2\Delta\om_{0,1}\,\om^{\G}_{g,n}\ .
  \label{eq:fixed-denominator-loop}
\end{equation}
The left hand side invites us to apply the {\it undeformed} topological recursion kernel $K_0$, whose denominator is built from $\om^0_{0,1}$, {\it i.e.,} the object we studied in Section~\ref{sec:kernel-origin}. So this means we can write the recursion instead as:
\begin{equation}
  \om^{\G}_{g,n}
  =\Res_{{\hat z}=0}K_0({\hat z}_1,{\hat z})
\left[\cQ^{\G}_{g,n}-2\Delta\om_{0,1}\,\om^{\G}_{g,n}\right]\ .
  \label{eq:fixed-kernel-TR}
\end{equation}
In other words, we can write the recursion in terms of the undeformed $K_0$, at the expense of shifting the object it acts on. Crucially,  all the Ramond parts of the disc $\Delta\omega_{0,1}$ are  explicitly included on the right hand side, as well as all the Ramond parts of $\omega_{0,2}$, which are included in the sum.
This is the topological recursion version of having the same NS type kernel, and moving all the R parts into the initial data, something that is built into ref.~\cite{norbury2024superweilpeterssonmeasuresmoduli}'s volume recursion approach.

So now if we take the  inverse Laplace transform of \eqref{eq:fixed-kernel-TR}, we will get  a  volume recursion with the usual \(D,R\), and with Ramond data in the initial data, what we seek for matching to ref.~\cite{norbury2024superweilpeterssonmeasuresmoduli}'s approach. But are  the recursions really  one and the same? If so, the \(s\)-deformed unstable inputs $\widehat F$ and $\widehat C$ that  ref.~\cite{norbury2024superweilpeterssonmeasuresmoduli} uses should be equivalent to the data in the $\Delta\om_{0,1}$ and $\Delta\om_{0,2}$ just defined here.

\medskip

\noindent Proving this occupy the next two sections.

\section{The disc proof}
\label{sec:disc-proof}
Let's look how ref.~\cite{norbury2024superweilpeterssonmeasuresmoduli} defines the initial unstable Ramond disc data. Define:
\begin{equation}
  {\widehat F}(\zh)
  :=
  \cL_b\!\left[\widehat V_{0,1}(s;b)\right](\zh)
  =
  \int_0^\infty e^{-\zh b}\,b\,\widehat V_{0,1}(s;b)\dd b\ .
  \label{eq:norbury-F-def}
\end{equation}
The disc recursion is:
\begin{eqnarray}
 \label{eq:norbury-disc-length-recursion} &&\hskip-0.6cm \widehat V_{0,1}(s;b)
  =
  \frac{s^2}{2}\\&&+
  \frac{1}{2b}
  \int_0^\infty\!\!\int_0^\infty
  xy\,D(b,x,y)\,
  \widehat V_{0,1}(s;x)\widehat V_{0,1}(s;y)
  \dd x\dd y \ . \nonumber
\end{eqnarray}
The first term results from the initial condition that the three-point function $V_{0,1}^{(2)}$ is unity.
Multiplying by  $b$ and applying $\cL_b$ this term transforms to:
\begin{equation}
  \cL_b\!\left[\frac{s^2}{2}\right](\zh)\equiv\oL_b\!\left[\frac{s^2}{2}b\right](\zh)
  =
  \frac{s^2}{2\zh^2}\ .
  \label{eq:disc-source-transform}
\end{equation}
For the next term, if we identify:
\begin{equation}
  P(x,y)
  =
  x\widehat V_{0,1}(s;x)\,
  y\widehat V_{0,1}(s;y)\ ,
  \label{eq:disc-P-def}
\end{equation}
we see that 
 $ \oL[P](\zh,\zh)
  =
  {\widehat F}(\zh)^2 $.
  \label{eq:disc-P-transform}
Using the $D$-kernel identity~\eqref{eq:D-transform-identity-from-kernel-origin} from earlier, the disc equation becomes~\cite{norbury2024superweilpeterssonmeasuresmoduli}:
\begin{equation}
  {\widehat F}(\zh)
  =
  \frac{s^2}{2\zh^2}
  +
  \frac12\,
  \PP_{\zh=0}\!\left[
  \frac{{\widehat F}(\zh)^2}{\cos(2\pi \zh)}
  \right]\ .
  \label{eq:norbury-F-PP-equation}
\end{equation}
What we need to show is that this defines precisely the same Ramond contributions to the disc  that  spectral curve~(\ref{eq:hard-edge-curve}) defines, {\it i.e.,} $\Delta\omega_{0,1}$. In fact, it was already shown in ref.~\cite{Johnson:2026jls} that the first several orders in the $s$ expansion match, but now we want to {\it prove} that they are the same. To state  what we need to prove, write:
\begin{equation}
  G({\hat z},E_0)=-2{\hat z} Y_{\G}({\hat z})\ ,
  \label{eq:G-def}
\end{equation} and define the  Ramond correction by subtracting off the purely NS part:
\begin{equation}
  W^R_{0,1}({\hat z},\G)=G({\hat z},E_0)-\cos(2\pi{\hat z})
  =-2{\hat z}\bigl(Y_{\G}({\hat z})-Y_0({\hat z})\bigr)\ .
  \label{eq:WJ-def}
\end{equation}
Then  the Ramond disc data are the same because of:

\bigskip
\noindent{\bf Theorem 1:}
\medskip

\noindent The function $W^R_{0,1}({\hat z},\G)$ satisfies:
\begin{equation}
  W^R_{0,1}({\hat z},\G)
  =\frac{\G^2}{2{\hat z}^2}
  -\PP_{{\hat z}=0}\left[\frac{W^R_{0,1}({\hat z},\G)^2}{2\cos(2\pi{\hat z})}\right].
  \label{eq:Johnson-disc-equation}
\end{equation}
We prove it as follows. Let's study $G({\hat z},E_0)$, which is:
\begin{eqnarray}
  &&G({\hat z},E_0)
  =\cos(2\pi{\hat z})-1
  \label{eq:G-Abel}\\
  &&\hskip1.5cm
  -{\hat z}\int_0^{E_0}\frac{f_0(u)}{\sqrt{{\hat z}^2+u}}\,\dd u
 +I_0(2\pi\sqrt{E_0})\frac{z}{{\hat z}}\ ,
  \nonumber 
\end{eqnarray}
(see~\eqref{eq:hard-edge-curve}), where $z^2{=}\zh^2{+}E_0$, and 
\begin{equation}
  f_0(u)=\frac{\dd}{\dd u}I_0(2\pi\sqrt u)\ .
  \label{eq:f0-def}
\end{equation}
At \(E_0{=}0\), the last term in \eqref{eq:G-Abel} is equal to unity, the penultimate term vanishes and we recover
\(G({\hat z},0){=}\cos(2\pi{\hat z})\).
With it, we establish:

\bigskip
\noindent{\bf Lemma 1:}
\medskip
\noindent As a formal power series in \(E_0\), \(G({\hat z},E_0)\) satisfies:
\begin{equation}
  \PP_{{\hat z}=0}\left[\frac{G({\hat z},E_0)^2}{2\cos(2\pi{\hat z})}\right]
  =\frac{\G^2}{2{\hat z}^2}\ .
  \label{eq:disc-projection-identity}
\end{equation}

To prove this, first differentiate \eqref{eq:G-Abel} with respect to~$E_0$.  Since \(z^2={\hat z}^2+E_0\), this gives:
\begin{eqnarray}
  \frac{\partial G}{\partial E_0}
  &=&-{\hat z}\frac{f_0(E_0)}{z}
  +f_0(E_0)\frac{z}{{\hat z}}
  +\frac{I_0(2\pi\sqrt{E_0})}{2{\hat z} z}\ ,
 \nonumber\\ 
&=&\frac{E_0f_0(E_0)+I_0(2\pi\sqrt{E_0})/2}{{\hat z} z}.
  \label{eq:G-derivative-mid}
\end{eqnarray}
On the other hand, differentiating the endpoint relation~\eqref{eq:endpoint-equation} and multiplying by $\sqrt{E_0}$ gives:
\begin{equation}
  \sqrt{E_0}\frac{\dd\G}{\dd E_0}
  =E_0f_0(E_0)+\frac{I_0(2\pi\sqrt{E_0})}{2}.
  \label{eq:Gamma-derivative}
\end{equation}
So we can write:
\begin{equation}
  \frac{\partial G}{\partial E_0}
  =\frac{\sqrt{E_0}\,(\dd\G/\dd E_0)}{{\hat z} z}\ .
  \label{eq:G-derivative}
\end{equation}
We will use this in a moment, but it is worthwhile pausing to note that this implies that \(W^R_{0,1}\) is itself a principal part. This is because $(z\zh)^{-1}$ is a power series in $E_0$ and in inverse powers of $\zh$. Its coefficient is a power series in $E_0$.
Integrating in \(E_0\) 
(and using that $G(\zh,0){=}\cos(2\pi\zh)$)) shows that
\begin{equation}
  W^R_{0,1}({\hat z},\G)=G({\hat z},E_0)-\cos(2\pi{\hat z})
  \label{eq:WJ-is-PP}
\end{equation}
is a principal part order by order in \(E_0\), (or in \(\G^2\)).

Now compute:
\begin{equation}
  \frac{\partial}{\partial E_0}
  \PP_{{\hat z}=0}\left[\frac{G^2}{2\cos(2\pi\zh)}\right]
   =\frac{\dd\G}{\dd E_0}\,
  \PP_{{\hat z}=0}\left[\frac{\sqrt{E_0}\,G}{{\hat z} z \cos(2\pi\zh)}\right].
  \label{eq:PP-derivative}
\end{equation}
In fact, the right hand side can be re-written, because:
\begin{equation}
 \frac{\sqrt{E_0}\,G}{z\cos(2\pi{\hat z})}
  =
  \frac{\widetilde\Gamma}{{\hat z}}+O({\hat z})\ ,
  \label{eq:auxiliary-PP}
\end{equation}
which is proven by using \eqref{eq:G-Abel} and subtracting \(\G/{\hat z}\):
\begin{widetext}    

\begin{align}
  \frac{\sqrt{E_0}\,G}{z\cos(2\pi\zh)}-\frac{\G}{{\hat z}}
  ={}&
  \frac{\sqrt{E_0}\,(\cos(2\pi\zh)-1)}{z\cos(2\pi\zh)}
  -\frac{\sqrt{E_0}\,{\hat z}}{z\cos(2\pi\zh)}
  \int_0^{E_0}\frac{f_0(u)}{\sqrt{{\hat z}^2+u}}\,\dd u
  +\frac{\G}{{\hat z}}\left(\frac{1}{\cos(2\pi\zh)}-1\right)\ ,
  \label{eq:auxiliary-regular}
\end{align}
\end{widetext}
and  seeing that the first term is
\(O({\hat z}^2)\), the middle term is \(O({\hat z})\), since the integral is
regular at \({\hat z}=0\), and the last term is \(O({\hat z})\).
Therefore we can write:
\begin{equation}
  \PP_{{\hat z}=0}
  \left[
  \frac{\sqrt{E_0}\,G}
  {{\hat z}z\cos(2\pi{\hat z})}
  \right]
  =
  \frac{\widetilde\Gamma}{{\hat z}^2}\ ,  
\end{equation}
and substituting this into \eqref{eq:PP-derivative} gives:
\begin{equation}
  \frac{\partial}{\partial E_0}
  \PP_{{\hat z}=0}\left[\frac{G^2}{2\cos(2\pi\zh)}\right]
  =\frac{\dd\G}{\dd E_0}\frac{\G}{{\hat z}^2}
  =\frac{\partial}{\partial E_0}\left[\frac{\G^2}{2{\hat z}^2}\right].
  \label{eq:PP-derivative-final}
\end{equation}
At \(E_0=0\), both sides of \eqref{eq:disc-projection-identity} vanish: the left side is
\(\PP[\cos(2\pi{\hat z})/2]=0\), and \(\G=0\).  This fixes the integration constant and proves {\bf Lemma 1}'s
\eqref{eq:disc-projection-identity}.

Now given the definition~(\ref{eq:WJ-def})
we have:
\begin{equation}
  \frac{G^2}{2\cos(2\pi\zh)}
  =\frac{\cos(2\pi\zh)}{2}+W^R_{0,1}+\frac{(W^R_{0,1})^2}{2\cos(2\pi\zh)}.
  \label{eq:G-square-expanded}
\end{equation}
The first term is regular, and \(W^R_{0,1}\) is a principal part, as  noticed above.  Taking principal parts and using
\eqref{eq:disc-projection-identity} proves {\bf Theorem 1}'s equation~(\ref{eq:Johnson-disc-equation}). 
As a non-linear equation, it defines a unique power series in $E_0$ (or $\G^2$) once the leading part is supplied, and so up to simple redefinition, the solutions are identical as power series.

Finally the redefinition is:
\begin{equation}
  W^R_{0,1}({\hat z},\G)=-\widehat F({\hat z},s),
  \qquad
  s^2=-\G^2\ .
  \label{eq:WJ-FN-identification}
\end{equation}
and we've established that indeed ref.~\cite{norbury2024superweilpeterssonmeasuresmoduli}'s initial unstable disc data defined by (\ref{eq:norbury-F-PP-equation}) is equivalent to the Ramond corrections $\Delta\omega_{0,1}$ encoded in ref.~\cite{Johnson:2026jls}'s random matrix model-derived  spectral curve.


Just for fun we can compare, from ref.~\cite{Johnson:2026jls}, (using the approach of studying solutions of a non-linear ODE (the ``string equation'')):
\begin{widetext}
\begin{align}
W_{0,1}^{\rm R}
={}&
\frac{\widetilde\Gamma^2}{2{\hat z}^2}
-\widetilde\Gamma^4
\left(
\frac{\pi^2}{4{\hat z}^2}
+\frac{1}{8{\hat z}^4}
\right)
+\widetilde\Gamma^6
\left(
\frac{11\pi^4}{24{\hat z}^2}
+\frac{\pi^2}{4{\hat z}^4}
+\frac{1}{16{\hat z}^6}
\right)
-\widetilde\Gamma^8
\left(
\frac{361\pi^6}{288{\hat z}^2}
+\frac{45\pi^4}{64{\hat z}^4}
+\frac{15\pi^2}{64{\hat z}^6}
+\frac{5}{128{\hat z}^8}
\right)
+O(\widetilde\Gamma^{10})\ ,
\label{eq:Ramond-disc-expansion}
\end{align} 
to the recursively obtained solution to equation~\eqref{eq:norbury-F-PP-equation}:
\begin{align}
\widehat F(\hat z)
={}&
\frac{s^2}{2\hat z^2}
+s^4\left(
\frac{\pi^2}{4\hat z^2}
+
\frac{1}{8\hat z^4}
\right)
+s^6\left(
\frac{11\pi^4}{24\hat z^2}
+
\frac{\pi^2}{4\hat z^4}
+
\frac{1}{16\hat z^6}
\right)
+s^8\left(
\frac{361\pi^6}{288\hat z^2}
+
\frac{45\pi^4}{64\hat z^4}
+
\frac{15\pi^2}{64\hat z^6}
+
\frac{5}{128\hat z^8}
\right)
+O(s^{10})\ .
\end{align}
On reflection, it is quite striking how the same information can be packaged quite so differently. Our proof above shows how precisely it comes about.

\section{The cylinder Proof}
\label{sec:cylinder-proof}

Now it is time to turn to ref.~\cite{norbury2024superweilpeterssonmeasuresmoduli}'s  cylinder initial data. Define:
\begin{equation}
  \widehat{\mathcal C}(\zh_1,\zh_2)=
  \int_0^\infty\!\int_0^\infty
  e^{-\zh_1b_1-\zh_2b_2}
  b_1b_2\hV_{0,2}(s;b_1,b_2)\dd b_1\dd b_2 .
  \label{eq:cylinder-laplace-def}
\end{equation}
The \((g,n)=(0,2)\) specialization of the volume  recursion~(\ref{eq:Norbury-volume-recursion}) is the
linear equation:
\begin{align}
  b_1\hV_{0,2}(s;b_1,b_2)
  ={}&\frac12\int_0^\infty\!\int_0^\infty xyD(b_1,x,y)
  \bigl[
  \hV_{0,1}(s;x)\hV_{0,2}(s;y,b_2)
  +\hV_{0,2}(s;x,b_2)\hV_{0,1}(s;y)
  \bigr]\dd x\dd y
  \nonumber\\
  &\hskip3cm+\int_0^\infty xR(b_1,b_2,x)\hV_{0,1}(s;x)\dd x\ .
  \label{eq:Norbury-02-length}
\end{align}
\end{widetext}
We have already worked out the rules for the Laplace transform of the $D$ kernel operating on objects, and recorded it in (\ref{eq:D-transform-identity-from-kernel-origin}).
Applying this to the two  terms in the first line of equation~\eqref{eq:Norbury-02-length} will contribute the following to $\widehat{\mathcal C}(\zh_1,\zh_2)$:
\begin{equation}
  \text{(first line)}\quad\Longrightarrow\quad\PP_{\zh_1=0}\left[\frac{\widehat F(\zh_1)}{\cos(2\pi\zh_1)}
  \widehat{\mathcal C}(\zh_1,\zh_2)\right]\ .
  \label{eq:D-part-02-transform}
\end{equation}
Turning to the second line with the \(R\)-kernel term, define the shorthand \(D(b,x){=}D(b,x,0)\) and notice that it is odd in $b$ (see equation~(\ref{eq:D-kernel})).
Denote its one-sided Laplace transform by:
\begin{equation}
  D_x(\zh)=\int_0^\infty e^{-\zh b}D(b,x)\dd b 
\ .
  \label{eq:dx-transform}
\end{equation} 
Now 
let's first examine  the double transform of the \(R\)-kernel before acting on anything, denoting it $R_x(\zh_1,\zh_2)$:
\begin{equation}
  R_x(\zh_1,\zh_2)=
  \int_0^\infty\!\int_0^\infty
  b_2 e^{-\zh_1b_1-\zh_2b_2}R(b_1,b_2,x)\dd b_1\dd b_2\ .
\end{equation}
Recall that:
\begin{equation}
R(b_1,b_2,x)=\frac12\left[D(b_1+b_2,x)+D(b_1-b_2,x)\right]\ ,
  \label{eq:R-kernel-standard-statement}
\end{equation}
so the  term with \(D(b_1-b_2,x)\), needs special care since its argument changes sign across the line \(b_1=b_2\).  For fixed \(b_2=b\), set \(u=b_1+b\) in the \(D(b_1+b,x)\) term, and set \(u=b_1-b\) in the \(D(b_1-b,x)\) term.  The latter change of variables gives \(u\in[-b,\infty)\).  Splitting the interval at \(u{=}0\), and then using \(u=-v\) on \([-b,0]\) together with the oddness \(D(-v,x){=}-D(v,x)\), the  \(b_1\)-integral yields:
\begin{widetext}
\begin{eqnarray}    
\int_0^\infty e^{-\zh_1 b_1}
\left[D(b_1+b,x)+D(b_1-b,x)\right]\dd b_1
=
2\cosh(\zh_1 b)D_x(\zh_1)
-
2\int_0^b
\cosh\!\bigl(\zh_1(b-u)\bigr)D(u,x)\dd u .
\label{eq:dx-oddness-intermediate}\nonumber
\end{eqnarray}
Multiplying this by \(b e^{-\zh_2 b}/2\), we next integrate over \(b\),  exchanging the order of integration in the second term. Using the integrals:
\begin{equation}
\int_0^\infty e^{-\zh_2 t}\cosh(\zh_1t)\dd t=
\frac12\left(\frac1{\zh_2-\zh_1}+\frac1{\zh_2+\zh_1}\right)\ ,
\end{equation}
and
\begin{equation}
\int_0^\infty t e^{-\zh_2 t}\cosh(\zh_1t)\dd t
=
{\cal B}(\zh_1,\zh_2)   \quad \text{where}\quad
  {\mathcal B}(\zh_1,\zh_2)
\equiv\frac12\left[\frac1{(\zh_1-\zh_2)^2}+\frac1{(\zh_1+\zh_2)^2}\right]\ ,
  \label{eq:Bsym-def}
\end{equation}
we get the following expression for the double transform:
\begin{align}
  R_x(\zh_1,\zh_2)={}&
  {\cal B}(\zh_1,\zh_2)[D_x(\zh_1)-D_x(\zh_2)]
  +\frac12
  \left(\frac1{\zh_1-\zh_2}-\frac1{\zh_1+\zh_2}\right)
  \int_0^\infty b e^{-\zh_2b}D(b,x)\dd b
  \ .
  \label{eq:R-double-transform-expanded}
\end{align}

So now we are ready to compute the transform of action of $R$ on an object. In~(\ref{eq:Norbury-02-length}), it is  $x$-integrated against the disc factor
\(x\hV_{0,1}(s;x)\dd x\). This will introduce the $\cal L$-Laplace transform $\widehat F(\zh)$ into the discussion.
Define, using 
 \eqref{eq:D-transform-identity-from-kernel-origin}:
\begin{equation}
H(\hat z)
=
\int_0^\infty x\,\widehat V_{0,1}(s;x)\,D_x(\hat z)\,dx
=
\operatorname{PP}_{\hat z=0}
\left[
\frac{\widehat F(\hat z)}{\cos(2\pi \hat z)}
\right]\ , \quad\text{so} \quad
H'(\hat z)=
-\int_0^\infty x\,\widehat V_{0,1}(s;x)
\left[
\int_0^\infty b\,e^{-\hat z b}D(b,x)\,db
\right]dx\ .
\end{equation} Then we see that the interaction transforms as:
\begin{align}
\int_0^\infty x\,\widehat V_{0,1}(s;x)\,
R_x(\hat z_1,\hat z_2)\,dx
={}&
\mathcal B(\hat z_1,\hat z_2)
\bigl[H(\hat z_1)-H(\hat z_2)\bigr]
-\frac12
\left(
\frac{1}{\hat z_1-\hat z_2}
-
\frac{1}{\hat z_1+\hat z_2}
\right)
H'(\hat z_2)=\operatorname{PP}_{\hat z_1=0}
\left[
\mathcal B(\hat z_1,\hat z_2)\,H(\hat z_1)
\right] \ ,
\end{align}
So  we have that the R-part of the cylinder equation~\eqref{eq:Norbury-02-length} is:\footnote{Note that Ref.~\cite{Norbury:2020vyi} computes a different expression for the action of $R$ in Lemma 6.11. However it is not weighted by $b_2$ as done here, and moreover it involves a projection onto the even sector. Differentiating by $\zh_2$ fixes the first difference, after multiplying by a minus sign, and then completing the resulting $(\zh_1-\zh_2)^{-2}$ into our ${\cal B}(\zh_1,\zh_2)$ takes care of the second, yielding agreement.}

\begin{equation}
\text{(second line)}\quad\Longrightarrow\quad\oL_{b_1,b_2}\left[
  \int_0^\infty b_2 xR(b_1,b_2,x)\hV_{0,1}(s;x)\dd x
  \right]
  =
  \PP_{\zh_1=0}\left[\frac{\widehat F(\zh_1)}{\cos(2\pi\zh_1)}
  {\mathcal B}(\zh_1,\zh_2)\right]\ .
  \label{eq:R-part-02-transform}
\end{equation}
Combining \eqref{eq:D-part-02-transform} and \eqref{eq:R-part-02-transform}, we see that ref.~\cite{norbury2024superweilpeterssonmeasuresmoduli}'s cylinder Ramond data $\widehat{\mathcal C}(\zh_1,\zh_2)$ is defined by this equation:
\begin{equation}
  \widehat{\mathcal C}(\zh_1,\zh_2)
  =
  \PP_{\zh_1=0}\left[
  \frac{\widehat F(\zh_1)}{\cos(2\pi\zh_1)}
  \left(\widehat{\mathcal C}(\zh_1,\zh_2)+{\mathcal B}(\zh_1,\zh_2)\right)
  \right]
  =
  \PP_{\zh_1=0}\left[
  \left(1-\frac{G(\zh_1,E_0)}{\cos(2\pi\zh_1)}\right)
  \left(\widehat{\mathcal C}(\zh_1,\zh_2)+{\mathcal B}(\zh_1,\zh_2)\right)
  \right].
  \label{eq:Norbury-cylinder-projection-G}
\end{equation}
where  equation \eqref{eq:WJ-FN-identification} was used, with~\eqref{eq:WJ-def}, to write in terms of the disc quantity $G(\zh,E_0)$ we studied in Section~\ref{sec:disc-proof}.
\end{widetext}

The goal now is to show that the cylinder data~(\ref{eq:pull-back-Bergman}) accompanying ref.~\cite{Johnson:2026jls}'s  spectral curve data    has the very same information. We want just the Ramond part, so we subtract  the Cauchy part and use \(z_i^2=\zh_i^2+E_0\):
\begin{eqnarray}
  \Delta B_{\G}(\zh_1,\zh_2)&=&B_{\G}(\zh_1,\zh_2)-B_0(\zh_1,\zh_2)\nonumber\\
&=&\dd_{\zh_1}\dd_{\zh_2}
  \log\left(
  \frac{\zh_1+\zh_2}{z_1+z_2}
  \right)\ .
  \label{eq:DeltaB}
\end{eqnarray}
So let us study the candidate quantity:
\begin{equation}
  \mathcal C(\zh_1,\zh_2)
  =
  \frac{\Delta B_{\G}(\zh_1,\zh_2)}{\dd\zh_1\dd\zh_2}
  =
  \frac{\zh_1\zh_2}{z_1z_2(z_1+z_2)^2}
  -\frac1{(\zh_1+\zh_2)^2}\ .
  \label{eq:CJ-two-forms}
\end{equation}

To complete our demonstration of the  equivalence between ref.~\cite{Johnson:2026jls}, spectral curve formulation and ref.~\cite{norbury2024superweilpeterssonmeasuresmoduli}'s formulation of the volume recursion, we will now prove:

\bigskip

\noindent{\bf Theorem 2:}
\bigskip

\noindent
The function $\mathcal C(\zh_1,\zh_2)$ satisfies the same equation~(\ref{eq:Norbury-cylinder-projection-G}) as ref.~\cite{norbury2024superweilpeterssonmeasuresmoduli}'s $\widehat{\mathcal C}(\zh_1,\zh_2)$, {\it i.e.}:
\begin{equation}
  \mathcal C(\zh_1,\zh_2)
  =
  \PP_{\zh_1=0}\left[
  \left(1-\frac{G(\zh_1,E_0)}{\cos(2\pi\zh_1)}\right)
  \left(\mathcal C+{\mathcal B}\right)(\zh_1,\zh_2)
  \right]\ ,
  \label{eq:CJ-solves-cylinder-equation}
\end{equation}
and furthermore  $\mathcal C=\widehat{\mathcal C}$ to all orders a power series in~$E_0$. 

To prove it, we add \eqref{eq:Bsym-def} to~(\ref{eq:CJ-two-forms}) to give:
\begin{equation}
  \left(\mathcal C+{\mathcal B}\right)(\zh_1,\zh_2)
  =
  \frac{\zh_1\zh_2(z_1^2+z_2^2)}
  {z_1z_2(\zh_1^2-\zh_2^2)^2}\ .
  \label{eq:CJ-plus-Bsym}
\end{equation}
and then we multiply:
\begin{align}
  \frac{G(\zh_1,E_0)}{\cos(2\pi\zh_1)}
  \left(\mathcal C+{\mathcal B}\right)(\zh_1,\zh_2)
  &={}
  \frac{G(\zh_1,E_0)}{\cos(2\pi\zh_1)}\frac{\zh_1}{z_1}
  \frac{\zh_2(z_1^2+z_2^2)}
  {z_2(\zh_1^2-\zh_2^2)^2}.
  \label{eq:G-times-cylinder-source}
\end{align}
Using  what we established in equation~(\ref{eq:auxiliary-PP}), we see that $\zh_1G(\zh_1,E_0)/[z_1\cos(2\pi\zh_1)]$ is regular and the remaining factor is harmless at $\zh_1{=}0$. 
 Therefore:
\begin{equation}
  \PP_{\zh_1=0}\left[
  \frac{G(\zh_1,E_0)}{\cos(2\pi\zh_1)}
  \left(\mathcal C+{\mathcal B}\right)(\zh_1,\zh_2)
  \right]=0\ .
  \label{eq:cylinder-projection-identity}
\end{equation}
 Furthermore, \({\mathcal B}\) is regular in \(\zh_1\), and so:
\begin{equation}
  \PP_{\zh_1=0}\left[\left(\mathcal C+{\mathcal B}\right)(\zh_1,\zh_2)\right]=\mathcal C(\zh_1,\zh_2) .
  \label{eq:CJ-PP-source}
\end{equation}
Combining \eqref{eq:cylinder-projection-identity} and \eqref{eq:CJ-PP-source} proves that $ \mathcal C$ solves   equation~(\ref{eq:CJ-solves-cylinder-equation}), of the same form as that solved by $ \widehat{\mathcal C}$, (\ref{eq:Norbury-cylinder-projection-G}). To finish proving the theorem, we can argue that the solution is unique as an  expansion in $E_0$ as follows:  Define:
\({ H}=\widehat{\mathcal C}-\mathcal C\).  Subtracting the two  equations gives:
\begin{equation}
  H=\PP_{\zh_1=0}\left[
  \left(1-\frac{G(\zh_1,E_0)}{\cos(2\pi\zh_1)}\right)H\right].
  \label{eq:H-uniqueness-equation}
\end{equation}
Recall that \(1-G/\cos(2\pi\zh_1)=O(E_0)\). So this means that  if \(H\) had a first nonzero coefficient at order \(E_0^r\), the right side of \eqref{eq:H-uniqueness-equation} would start at higher order.  Hence that coefficient vanishes.  Reasoning by induction to all orders shows that  \(H=0\), establishing {\bf Theorem 2}. 

Therefore, translating we have:
\begin{equation}
  \left.{\widehat{\cal C}}(\zh_1,\zh_2)\right|_{s^2=-\G^2}
  =
  \frac{\Delta B_{\G}(\zh_1,\zh_2)}{\dd\zh_1\dd\zh_2}\ .
  \label{eq:cylinder-identification}
\end{equation}

Just to be explicit, expanding \eqref{eq:DeltaB} gives:
\begin{widetext}
\begin{align}
\mathcal C(\zh_1,\zh_2)\equiv  \frac{\Delta B_{\G}}{\dd\zh_1\dd\zh_2}
  ={}&-\frac{\G^2}{2\zh_1^2\zh_2^2}
  +\G^4\left[\frac{\pi^2}{\zh_1^2\zh_2^2}
  +\frac{3}{8}\left(\frac{1}{\zh_1^4\zh_2^2}+\frac{1}{\zh_1^2\zh_2^4}\right)\right]
  \nonumber\\
  &\hskip1.0cm-\G^6\Bigg[
  \frac{13\pi^4}{4\zh_1^2\zh_2^2}
  +\frac{3\pi^2}{2}\left(\frac{1}{\zh_1^4\zh_2^2}+\frac{1}{\zh_1^2\zh_2^4}\right)
  +\frac{5}{16}\left(\frac{1}{\zh_1^6\zh_2^2}+\frac{1}{\zh_1^2\zh_2^6}\right)
  +\frac{3}{8}\frac{1}{\zh_1^4\zh_2^4}
  \Bigg]+O(\G^8)\ .
  \label{eq:DeltaB-expansion}
\end{align}
while recursively solving our equation~(\ref{eq:Norbury-cylinder-projection-G}) expressing Norbury's cylinder Ramond data gives:
\begin{align}
\widehat{\mathcal C}(\hat z_1,\hat z_2)
={}&
\frac{s^2}{2\hat z_1^2\hat z_2^2}
+s^4\left[
\frac{\pi^2}{\hat z_1^2\hat z_2^2}
+\frac{3}{8}\left(
\frac{1}{\hat z_1^4\hat z_2^2}
+
\frac{1}{\hat z_1^2\hat z_2^4}
\right)
\right]
\nonumber\\
&\hskip2.5cm+s^6\left[
\frac{13\pi^4}{4\hat z_1^2\hat z_2^2}
+\frac{3\pi^2}{2}\left(
\frac{1}{\hat z_1^4\hat z_2^2}
+
\frac{1}{\hat z_1^2\hat z_2^4}
\right)
+\frac{5}{16}\left(
\frac{1}{\hat z_1^6\hat z_2^2}
+
\frac{1}{\hat z_1^2\hat z_2^6}
\right)
+\frac{3}{8}\frac{1}{\hat z_1^4\hat z_2^4}
\right]
+O(s^8)\ .
\label{eq:C-expansion}
\end{align}
which is equivalent to (after Laplace transforming according to 
 convention \eqref{eq:laplace-convention}):
\begin{align}
  \hV_{0,2}(s;b_1,b_2)
  ={}&\frac{s^2}{2}
  +s^4\left[\pi^2+\frac{b_1^2+b_2^2}{16}\right]
  +s^6\left[
  \frac{13\pi^4}{4}+\frac{\pi^2}{4}(b_1^2+b_2^2)
  +\frac{b_1^4+b_2^4}{384}+\frac{b_1^2b_2^2}{96}
  \right]+O(s^8)\ ,
  \label{eq:cylinder-volume-expansion}
\end{align}
\end{widetext}
as can be obtained by using the volume recursion directly.
The core point is that~\eqref{eq:C-expansion} and~\eqref{eq:DeltaB-expansion} are the same since $s^2=-\G^2$. Just as remarked at the end of the previous section, it is remarkable that this same series is packaged, as physical information in two such different ways. 
 
 \section{Concluding Remarks}
In conclusion we've shown that topological recursion applied to the spectral curve  of ref.~\cite{Johnson:2026jls} is precisely equivalent to the full direct volume recursion, including both NS and R sectors, where (as pointed out in ref.~\cite{norbury2024superweilpeterssonmeasuresmoduli}) the structure  and kernels $D$ and $R$ are the same as for the pure NS case~\cite{Stanford:2019vob,Norbury:2020vyi}, with the Ramond data entering through extra unstable (disc and cylinder) contributions. The key steps were to show how to correspondingly rewrite topological recursion for the curve in a way that used the purely NS kernel, and then to prove that the unstable data on both sides were precisely equivalent. In other words, the disc data of ref.~\cite{norbury2024superweilpeterssonmeasuresmoduli} turned out to contain the same information as that specified by the curve (and hence the underlying random matrix model presented in ref.~\cite{Johnson:2026jls}, while the cylinder data is equivalent to ref.~\cite{Johnson:2026jls}'s observation that topological recursion must use a particular non-Cauchy form for  the Bergman kernel.

Strikingly   there is a different choice to be made, however (as discussed in Section~\ref{sec:alternatives}). The spectral curve naturally produces a  kernel $K$ that is {\it different} from the undeformed one, $K_0$. Direct Laplace transform implies a new $D$ and $R$ in an alternative (but evidently equivalent) volume recursion. While the $D$ and $R$ seem hard to write in closed form in length variables, they clearly exist, and (presumably just like the undeformed $D$ and~$R$) have a geometrical meaning.  Whether they can be interpreted in terms of properties of classes of geodesics on Ramond-deformed pairs of pants would be interesting to explore in a future project.

\begin{acknowledgments}
CVJ   thanks  the  US Department of Energy (under grant \protect{DE-SC}~0011702)  for  support,  and  Amelia for her support and patience.    
\end{acknowledgments}

\bibliographystyle{apsrev4-1}
\bibliography{references}


\end{document}